\documentclass[twocolumn,description,aps]{revtex4-1}
\usepackage{graphicx,epstopdf,amsmath,amssymb,multirow,CJK,color}
\usepackage[german,english]{babel}
\usepackage[colorlinks, linkcolor=blue,anchorcolor=blue,citecolor=blue,urlcolor=blue]{hyperref}
\usepackage{booktabs}

\begin{document}

\title{Intrinsic ferromagnetic and antiferromagnetic axion insulators in van der Waals materials Mn\emph{X}$_{2}$\emph{B}$_{2}$\emph{T}$_{6}$ family}

\author{Yan Gao}
\author{Kai Liu}\email{kliu@ruc.edu.cn}
\author{Zhong-Yi Lu}\email{zlu@ruc.edu.cn}

\affiliation{Department of Physics and Beijing Key Laboratory of Opto-electronic Functional Materials $\&$ Micro-nano Devices, Renmin University of China, Beijing 100872, China}

\date{\today}

\begin{abstract}
The MnBi$_{2}$Te$_{4}$ family has attracted significant attention due to its rich topological states such as the quantum anomalous Hall (QAH) insulator state, the axion insulator state, and the magnetic Weyl semimetal state. Nevertheless, the intrinsic antiferromagnetic (AFM) interlayer coupling in MnBi$_{2}$Te$_{4}$ partly hinders the realization of ``high-temperature'' QAH effect. Here, by using first-principles electronic structure calculations, we design a new class of materials Mn\emph{X}$_{2}$\emph{B}$_{2}$\emph{T}$_{6}$ (\emph{X}=Ge, Sn, or Pb; \emph{B}=Sb or Bi; \emph{T}=Se or Te) based on the \emph{X}$_{2}$\emph{B}$_{2}$\emph{T}$_{5}$ structures rather than the Bi$_{2}$Te$_{3}$ family. We find that each septuple-layer Mn\emph{B}$_{2}$\emph{T}$_{4}$ is sandwiched by two [\emph{X}\emph{T}] layers, which may turn the AFM interlayer coupling into a ferromagnetic (FM) coupling. The calculations specifically demonstrate that \emph{MnGe}$_{2}$\emph{Sb}$_{2}$\emph{Te}$_{6}$, \emph{MnGe}$_{2}$\emph{Bi}$_{2}$\emph{Te}$_{6}$, and \emph{MnPb}$_{2}$\emph{Bi}$_{2}$\emph{Te}$_{6}$ are FM axion insulators, while MnGe$_{2}$Sb$_{2}$Se$_{6}$, MnGe$_{2}$Bi$_{2}$Se$_{6}$, MnSn$_{2}$Sb$_{2}$Te$_{6}$, and MnSn$_{2}$Bi$_{2}$Te$_{6}$ are A-type AFM axion insulators. These seven materials all have an out-of-plane easy axis of magnetization. The Mn\emph{X}$_{2}$\emph{B}$_{2}$\emph{T}$_{6}$ family thus offers a promising platform beyond the MnBi$_{2}$Te$_{4}$ family for the realization of quantized magnetoelectric effect and ``high-temperature'' QAH effect in future experiments.
\end{abstract}

\date{\today} \maketitle

\begin{figure*}[!th]
	\centering
	\includegraphics[width=0.72\textwidth]{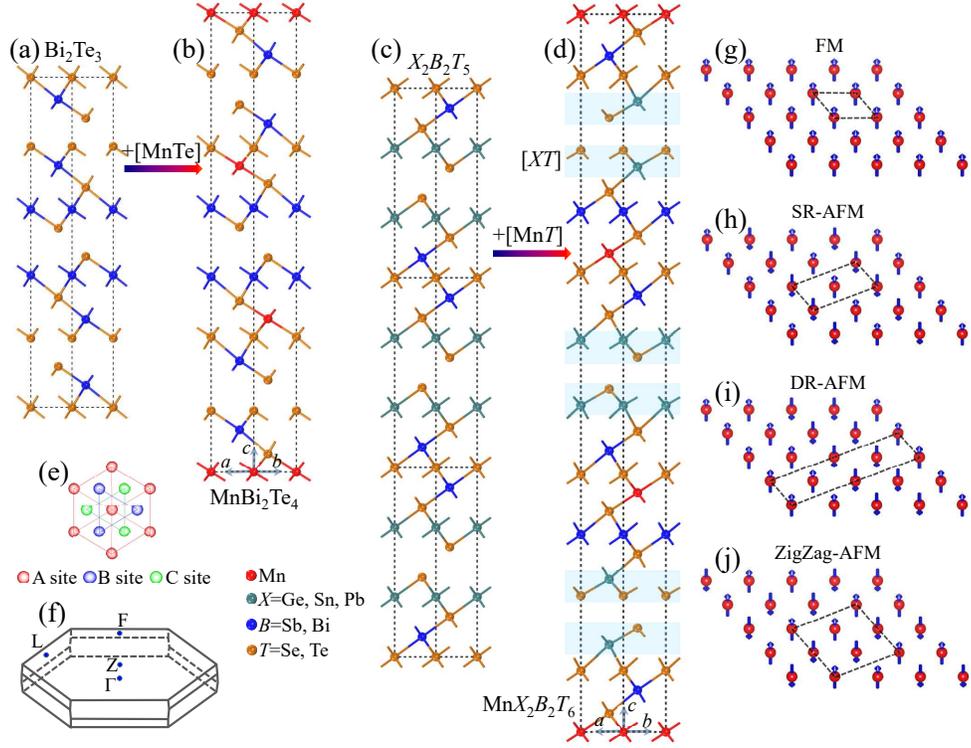}
	\caption{(Color online) Optimized crystal structures and typical magnetic configurations. Crystal structures of (a) Bi$_{2}$Te$_{3}$, (b) MnBi$_{2}$Te$_{4}$, (c) \emph{X}$_{2}$\emph{B}$_{2}$\emph{T}$_{5}$ (\emph{X}=Ge, Sn, or Pb; \emph{B}=Sb or Bi; \emph{T}=Se or Te), and (d) Mn\emph{X}$_{2}$\emph{B}$_{2}$\emph{T}$_{6}$. (e) Schematic of the top view along the \emph{c}-axis direction. (f) Brillouin zone (BZ) of the primitive cell of Mn\emph{X}$_{2}$\emph{B}$_{2}$\emph{T}$_{6}$. The $\Gamma$ (0,0,0), L (0,$\pi$,0), F ($\pi$,$\pi$,0), and Z ($\pi$,$\pi$,$\pi$) are four inequivalent inversion-invariant momenta points. Four possible magnetic configurations on a triangular lattice formed by intralayer Mn atoms, namely (g) ferromagnetic (FM), (h) single row antiferromagnetic (SR-AFM), (i) double row AFM (DR-AFM), and (j) ZigZag AFM (ZigZag-AFM) states. The black dotted lines in (a-d) and (g-j) represent crystal cells and magnetic unit cells, respectively.}
	\label{fig_structure}
\end{figure*}

$Introduction$: In recent years, MnBi$_{2}$Te$_{4}$ as an intrinsic antiferromagnetic (AFM) topological insulator (TI) has attracted intensive attention~\cite{1Otrokov,2Gong,3Deng,4Liu,5Zhang,6Li,7Zeugner,8Hao,9Chen,10Li,11Zhang,12LiYan}, because its unique structural characteristics endow with rich magnetic topological states, including the axion insulators (AXIs), the quantum anomalous Hall (QAH) insulators, and the magnetic Weyl semimetals. The tetradymite-type MnBi$_{2}$Te$_{4}$~\cite{13Lee} is composed of Te-Bi-Te-Mn-Te-Bi-Te septuple-layer (SL) blocks stacking along the \emph{c} axis via the van der Waals (vdW) interaction, which can be regarded as an [MnTe] layer being inserted into the middle of each Bi$_{2}$Te$_{3}$ unit with a Te-Bi-Te-Bi-Te quintuple layer (QL), as shown in Figs.~\ref{fig_structure}(a) and~\ref{fig_structure}(b). The ferromagnetic (FM) coupling between intralayer Mn atoms and the AFM coupling between interlayer Mn atoms result in an A-type AFM configuration in MnBi$_{2}$Te$_{4}$, whose easy magnetic axis is along the out-of-plane direction~\cite{2Gong,5Zhang,6Li}. Obviously, the topological properties of MnBi$_{2}$Te$_{4}$ derive from the parent TI Bi$_{2}$Te$_{3}$, while the magnetism originates from Mn$^{2+}$ ions in the [MnTe] layer. On one hand, the intrinsic topological properties of MnBi$_{2}$Te$_{4}$-related ternary chalcogenides (such as MnSb$_{2}$Te$_{4}$, MnSb$_{2}$Se$_{4}$, and MnBi$_{2}$Se$_{4}$) are very sensitive to the strength of spin-orbit coupling (SOC)~\cite{5Zhang,14LiYu,15Zhang,16Zhou}, since the common origin of TIs is the SOC-induced band inversion~\cite{17Qi,18Hasan}. On the other hand, due to the intrinsic A-type AFM magnetism in bulk MnBi$_{2}$Te$_{4}$, the zero-field QAH effect appears only in an odd-number SL films~\cite{3Deng,6Li}. For example, it was reported that the zero-field QAH effect was observed at 1.4~K in a five-SL MnBi$_{2}$Te$_{4}$ film, and an external magnetic field that enforces the interlayer coupling from AFM to FM further raised the quantization temperature to 6.5~K or even 45~K~\cite{3Deng,19Ge}. Therefore, searching for stoichiometric MnBi$_{2}$Te$_{4}$-related compounds satisfying both the TI properties with strong SOC and the FM ground state with relatively strong interlayer coupling has become one of the main subjects for the realization of ``high-temperature'' QAH effect.

At present, much effort has been devoted to realizing an interlayer FM coupling between the MnBi$_{2}$Te$_{4}$-related layers, such as charge$/$element doping~\cite{20Han,21Hou}, constructing heterojunctions with different \emph{d}-orbital vdW materials~\cite{22Li,23Zhu,24Fu}, and inserting Bi$_{2}$Te$_{3}$ or Sb$_{2}$Te$_{3}$ layers in between the MnBi$_{2}$Te$_{4}$ layers to increase the Mn-Mn interlayer distance~\cite{25Hu,26Lu,27Qi}. We notice that all these proposals are implemented in the framework of adopting the TI Bi$_{2}$Te$_{3}$ family as parent materials. We would like to ask whether there are other TIs as parent materials beyond the Bi$_{2}$Te$_{3}$ family that can achieve the intrinsic magnetic TIs with an FM coupling.

In this study, we find that the TI materials \emph{X}$_{2}$\emph{B}$_{2}$\emph{T}$_{5}$ (\emph{X}=Ge, Sn, or Pb; \emph{B}=Sb or Bi; \emph{T}=Se or Te) [see Fig.~\ref{fig_structure}(c)], including the TIs Ge$_{2}$Sb$_{2}$Te$_{5}$ and Pb$_{2}$Bi$_{2}$Te$_{5}$ that were synthesized experimentally and predicted theoretically~\cite{28Petrov,29Kim,30Petrov,31Silkin}, can serve as the parent materials of Mn\emph{X}$_{2}$\emph{B}$_{2}$\emph{T}$_{6}$ [Fig.~\ref{fig_structure}(d)]. The Mn\emph{X}$_{2}$\emph{B}$_{2}$\emph{T}$_{6}$ can be considered as intercalating an [Mn\emph{T}] layer into the middle of each \emph{X}$_{2}$\emph{B}$_{2}$\emph{T}$_{5}$ nonuple-layer block [Figs.~\ref{fig_structure}(c) and~\ref{fig_structure}(d)], which is similar to the case of Bi$_{2}$Te$_{3}$ serving as the parent material of MnBi$_{2}$Te$_{4}$~\cite{13Lee}. One common feature of Bi$_{2}$Te$_{3}$, MnBi$_{2}$Te$_{4}$, \emph{X}$_{2}$\emph{B}$_{2}$\emph{T}$_{5}$ and Mn\emph{X}$_{2}$\emph{B}$_{2}$\emph{T}$_{6}$ is that each atomic layer in these compounds forms a triangular lattice with only three distinct positions labeled as A, B, and C [see Fig.~\ref{fig_structure}(e)]. Furthermore, Mn\emph{X}$_{2}$\emph{B}$_{2}$\emph{T}$_{6}$ can also be viewed as each SL Mn\emph{B}$_{2}$\emph{T}$_{4}$ sandwiched by two [\emph{XT}] layers. It is also similar to the experimentally prepared Pb$_{2}$Bi$_{2}$Te$_{5}$~\cite{30Petrov}, which is regarded as each Bi$_{2}$Te$_{3}$ QL sandwiched by two [PbTe] layers [see Figs.~\ref{fig_structure}(a) and~\ref{fig_structure}(c)].

We have systematically studied the electronic, magnetic, and topological properties of the Mn\emph{X}$_{2}$\emph{B}$_{2}$\emph{T}$_{6}$ family materials by using first-principles electronic structure calculations. The computational details are provided in the Supplemental Material (SM)~\cite{39SM}, which includes Refs.~\cite{32PBE,33Kresse,34Perdew,35Grimme,36Monkhorst,37Togo,38Wu}.



$Results$: To evaluate the stability of Mn\emph{X}$_{2}$\emph{B}$_{2}$\emph{T}$_{6}$ family (\emph{X}=Ge, Sn, or Pb; \emph{B}=Sb or Bi; \emph{T}=Se or Te), we have first calculated the phonon spectra of the Mn\emph{X}$_{2}$\emph{B}$_{2}$\emph{T}$_{6}$ monolayers, in which seven materials MnGe$_{2}$Sb$_{2}$Se$_{6}$, \emph{MnGe}$_{2}$\emph{Sb}$_{2}$\emph{Te}$_{6}$, MnGe$_{2}$Bi$_{2}$Se$_{6}$, \emph{MnGe}$_{2}$\emph{Bi}$_{2}$\emph{Te}$_{6}$, MnSn$_{2}$Sb$_{2}$Te$_{6}$, MnSn$_{2}$Bi$_{2}$Te$_{6}$, and \emph{MnPb}$_{2}$\emph{Bi}$_{2}$\emph{Te}$_{6}$ monolayers are found to have no soft modes across the Brillouin zone (BZ), indicating their dynamical stabilities. In comparison, the other five materials have imaginary phonon modes (see Fig. S1 in the SM~\cite{39SM}). Thus, we focus on the seven materials with dynamical stabilities.

We now consider the various magnetic configurations in the monolayer forms of these seven materials and then examine the interlayer magnetic coupling to identify their magnetic ground states. Since the Mn atoms form a triangular lattice in each plane, we consider four typical magnetic structures [see Figs.~\ref{fig_structure}(g)-~\ref{fig_structure}(j)], namely FM, single row AFM (SR-AFM), double row AFM (DR-AFM), and ZigZag AFM (ZigZag-AFM) states. Our calculations show that the FM state has the lowest energy among these magnetic states (see Table SI in the SM~\cite{39SM}). Note that the energy is not available for the ZigZag-AFM state, for which the calculations tend to be not converged or converged to the nonmagnetic (NM) state, suggesting that this magnetic structure is unstable. Meanwhile, we have tested different Hubbard $U$ values ($U$=3, 4, and 5~eV) on the Mn 3\emph{d} orbitals, and found that the selection of $U$ values has almost no effect on the structural and magnetic properties. Thus, in the following discussion we take $U=4$~eV, which is the same as those in Refs.~\cite{6Li,14LiYu}. From Table SI, we see that the bond angles of Mn-\emph{T}-Mn in these seven materials are in the range of 92.1$^{\circ}$-94.8$^{\circ}$ (close to 90$^{\circ}$), hence the FM coupling between the intralayer Mn atoms may be understood from the Goodenough-Kanamori rule~\cite{40Goodenough,41Kanamori,42Goodenough}. Furthermore, the calculated magnetic anisotropy energies (MAE) show that their easy magnetization axes all tend to be out-of-plane and the MAE values are comparable to the one of MnBi$_{2}$Te$_{4}$ (0.2~meV)~\cite{6Li}. Thus these seven materials all exhibit the intralayer FM coupling with an out-of-plane easy magnetic axis.

\begin{table*}[!th]
\caption{\label{tab:I} The structural, magnetic, and topological properties of bulk compounds MnGe$_{2}$Sb$_{2}$Se$_{6}$, \emph{MnGe}$_{2}$\emph{Sb}$_{2}$\emph{Te}$_{6}$, MnGe$_{2}$Bi$_{2}$Se$_{6}$, \emph{MnGe}$_{2}$\emph{Bi}$_{2}$\emph{Te}$_{6}$, MnSn$_{2}$Sb$_{2}$Te$_{6}$, MnSn$_{2}$Bi$_{2}$Te$_{6}$, and \emph{MnPb}$_{2}$\emph{Bi}$_{2}$\emph{Te}$_{6}$, including the lattice constants (\emph{a} and \emph{c}) optimized in their respective magnetic ground states, the bond angles of Mn-\emph{T}-Mn between the two nearest Mn atoms, the atomic magnetic moments (\emph{M}$_{\rm{atom}}$), the energy differences of the A-type AFM state with respect to the FM state ($\Delta{E}_{\rm{AFM-FM}}$), the numbers of occupied bands with even and odd parities ($n_{occ}^+$,$n_{occ}^-$) at the eight inversion-invariant momenta points ($\Lambda_{\alpha}$) with the inclusion of SOC, and the Z$_{4}$ invariants.}
\begin{center}
\begin{tabular*}{2.05\columnwidth}{@{\extracolsep{\fill}}ccccccccccc}
\hline\hline
\multirow{2}{*}{Structures} & \multicolumn{2}{c}{Lattice constants} & {Angles ($^{\circ}$)} & {\emph{M}$_{\rm{atom}}$}  & {$\Delta{E}_{\rm{AFM-FM}}$}  & \multicolumn{4}{c}{$\Lambda_{\alpha}$} & \multirow{2}{*}{Z$_{4}$} \\
\cline{2-3}\cline{7-10}
  & {\emph{a} (\text{\AA})} & {\emph{c} (\text{\AA})} & {Mn-\emph{T}-Mn} & ($\mu_{\rm{B}}$) & (meV/Mn) & $\Gamma$ (0,0,0) & 3$\times$L (0,$\pi$,0) & 3$\times$F ($\pi$,$\pi$,0) & Z ($\pi$,$\pi$,$\pi$) &  \\
\hline
 MnGe$_{2}$Sb$_{2}$Se$_{6}$ & 3.98 & 60.16 & 93.1 & 4.55 & -0.048 & (88,86) & (87,87) & (86,88) & (87,87) & 2 \\
 \emph{MnGe}$_{2}$\emph{Sb}$_{2}$\emph{Te}$_{6}$ & 4.26 & 62.59 & 92.8 & 4.52 & 0.198 & (45,42) & (43,44) & (43,44) & (43,44) & 2 \\
 MnGe$_{2}$Bi$_{2}$Se$_{6}$ & 4.03 & 60.39 & 94.1 & 4.55 & -0.035 & (108,106) & (107,107) & (106,108) & (107,107) & 2 \\
 \emph{MnGe}$_{2}$\emph{Bi}$_{2}$\emph{Te}$_{6}$ & 4.30 & 62.93 & 93.6 & 4.53 & 0.342 & (55,52) & (53,54) & (53,54) & (53,54) & 2 \\
 MnSn$_{2}$Sb$_{2}$Te$_{6}$ & 4.34 & 63.43 & 94.4 & 4.54 & -0.042 & (88,86) & (87,87) & (86,88) & (87,87) & 2 \\
 MnSn$_{2}$Bi$_{2}$Te$_{6}$ & 4.39 & 64.07 & 95.3 & 4.55 & -0.023 & (108,106) & (107,107) & (106,108) & (107,107) & 2 \\
 \emph{MnPb}$_{2}$\emph{Bi}$_{2}$\emph{Te}$_{6}$ & 4.43 & 65.60 & 95.8 & 4.55 & 0.190 & (55,52) & (53,54) & (53,54) & (53,54) & 2 \\
\hline\hline
\end{tabular*}
\end{center}
\end{table*}

\begin{figure}[th]
	\centering
	\includegraphics[width=0.46\textwidth]{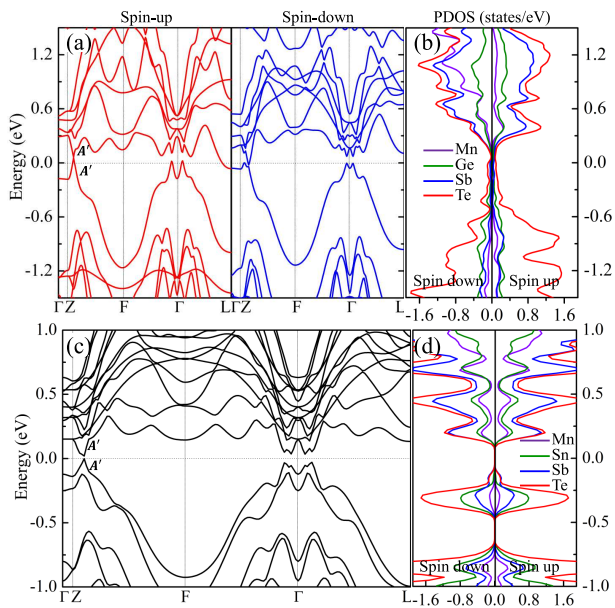}
	\caption{(Color online) (a) Spin-polarized band structures and (b) projected density of states (PDOS) of the bulk \emph{MnGe}$_{2}$\emph{Sb}$_{2}$\emph{Te}$_{6}$ in the FM ground state. (c) Band structure and (d) PDOS of the bulk MnSn$_{2}$Sb$_{2}$Te$_{6}$ in the A-type AFM ground state. The SOC is not included in the calculations.}
	\label{fig_spinband}
\end{figure}


The tetradymite-type Mn\emph{X}$_{2}$\emph{B}$_{2}$\emph{T}$_{6}$ can crystalize in a rhombohedral structure with the space group \emph{D}$_{3d}^5$ (No. 166), which is composed of eleven-layer blocks stacking along the \emph{c}-axis through vdW interaction [see Fig.~\ref{fig_structure}(d)]. We now only need to consider the interlayer magnetic coupling to identify the magnetic ground states for their bulk materials. From the calculated results in Table \ref{tab:I}, we find that for bulk compounds MnGe$_{2}$Sb$_{2}$Se$_{6}$, MnGe$_{2}$Bi$_{2}$Se$_{6}$, MnSn$_{2}$Sb$_{2}$Te$_{6}$ and MnSn$_{2}$Bi$_{2}$Te$_{6}$ the interlayer AFM states are energetically lower than their respective FM states, indicating that their magnetic ground states are in the A-type AFM configuration with the intralayer FM and interlayer AFM couplings, which are the same as the ones of MnBi$_{2}$Te$_{4}$~\cite{2Gong,5Zhang,6Li}. It is worth noting that the energy differences between their A-type AFM and FM states are very small and comparable to that in MnBi$_{6}$Te$_{10}$~\cite{43Klimovskikh}, which is regarded as the boundary between the A-type AFM and FM ground states in the MnBi$_{2n}$Te$_{3n+1}$ family. Thus, a small external magnetic field can force the system to become ferromagnetic. On the other hand, the bulk compounds \emph{MnGe}$_{2}$\emph{Sb}$_{2}$\emph{Te}$_{6}$, \emph{MnGe}$_{2}$\emph{Bi}$_{2}$\emph{Te}$_{6}$, and \emph{MnPb}$_{2}$\emph{Bi}$_{2}$\emph{Te}$_{6}$ prefer the FM states, which are lower in energy than the respective A-type AFM states by 0.19-0.34~meV/Mn. On the contrary, the energy of the A-type AFM ground state of MnBi$_{2}$Te$_{4}$~\cite{14LiYu} is 0.3~meV/Mn lower than that of its FM state. Thus, the bulk compounds \emph{MnGe}$_{2}$\emph{Sb}$_{2}$\emph{Te}$_{6}$, \emph{MnGe}$_{2}$\emph{Bi}$_{2}$\emph{Te}$_{6}$, and \emph{MnPb}$_{2}$\emph{Bi}$_{2}$\emph{Te}$_{6}$ host the FM ground states with an out-of-plane easy axis of magnetization, which are desirable for the realization of intrinsic ``high-temperature'' QAH effect.

\begin{figure*}[!th]
	\centering
	\includegraphics[width=0.82\textwidth]{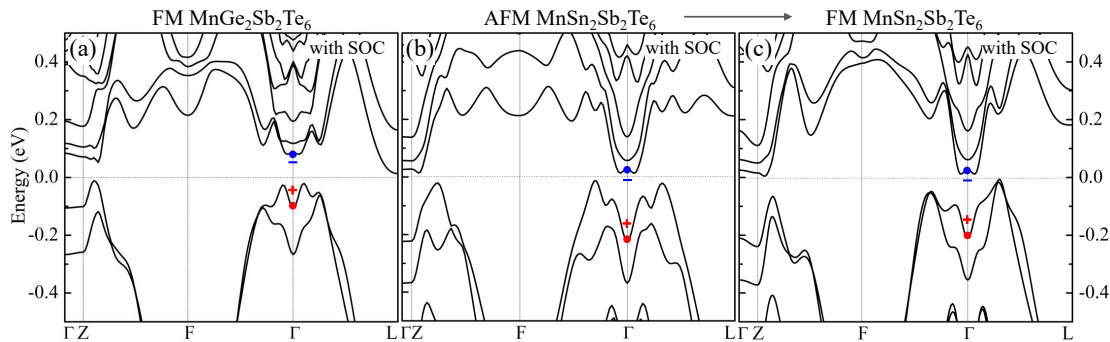}
	\caption{(Color online) Band structures of the (a) FM \emph{MnGe}$_{2}$\emph{Sb}$_{2}$\emph{Te}$_{6}$ bulk, (b) A-type AFM MnSn$_{2}$Sb$_{2}$Te$_{6}$ bulk, and (c) FM MnSn$_{2}$Sb$_{2}$Te$_{6}$ bulk along the high-symmetry paths in the BZ calculated with the SOC. The even and odd parities of the valence band maximum and the conduction band minimum at the $\Gamma$ point are labeled by ``+'' and ``$-$'', respectively.}
	\label{fig_SOCband}
\end{figure*}

Next, we study the electronic properties of these seven materials. Considering that the \emph{X}, \emph{B}, and \emph{T} in Mn\emph{X}$_{2}$\emph{B}$_{2}$\emph{T}$_{6}$ respectively represent the same column in the periodic table and have the same valence, they should have similar electronic properties with the same magnetic structures. Here the FM \emph{MnGe}$_{2}$\emph{Sb}$_{2}$\emph{Te}$_{6}$ and A-type AFM MnSn$_{2}$Sb$_{2}$Te$_{6}$ are selected as examples for illustration, and the electronic properties of other materials can be obtained in the SM~\cite{39SM}.

The calculated band structure and projected density of state (PDOS) of the FM \emph{MnGe}$_{2}$\emph{Sb}$_{2}$\emph{Te}$_{6}$ are respectively shown in Figs.~\ref{fig_spinband}(a) and~\ref{fig_spinband}(b). We see that it exhibits metallic characteristics and the bands around the Fermi level are mainly contributed by the \emph{p} orbitals of Te, Sb, and Ge atoms. In contrast, for the AFM MnSn$_{2}$Sb$_{2}$Te$_{6}$, from the calculated band structure [Fig.~\ref{fig_spinband}(c)] and PDOS [Fig. ~\ref{fig_spinband}(d)] we see that it exhibits an insulating behavior and the bands around the Fermi level are also mainly contributed by the \emph{p} orbitals of Te, Sb, and Sn atoms. The further analysis shows that both the FM \emph{MnGe}$_{2}$\emph{Sb}$_{2}$\emph{Te}$_{6}$ and AFM MnSn$_{2}$Sb$_{2}$Te$_{6}$ have the same little point group of $\emph{C}_{s}$ along the Z-F path, and the two bands that participate in the crossing around the Fermi level belong to the same one-dimensional irreducible representations of $A^{\prime}$ [see Figs.~\ref{fig_spinband}(a) and~\ref{fig_spinband}(c)]. Thus, several meV gaps open in these band crossings.

When the SOC is included, the band structures display substantial changes. Figures~\ref{fig_SOCband}(a) and~\ref{fig_SOCband}(b) show that the valence bands and the conduction bands of both FM \emph{MnGe}$_{2}$\emph{Sb}$_{2}$\emph{Te}$_{6}$ and A-type AFM MnSn$_{2}$Sb$_{2}$Te$_{6}$ are inverted at the $\Gamma$ point around the Fermi level, which implies that they may have nontrivial topological properties. We would like to remind that for centrosymmetric three-dimensional insulators without time-reversal symmetry, the quantized axion angle $\theta=0$ or $\pi$ corresponds to the invariant Z$_{4}$=0 (topological trivial) or 2 (topological nontrivial), respectively~\cite{44Xu,45Turner}. Because of the presence of the inversion (\emph{I}) symmetry in Mn\emph{X}$_{2}$\emph{B}$_{2}$\emph{T}$_{6}$ family, the parity-based higher-order Z$_{4}$ invariant can be adopted to characterize the topological properties of Mn\emph{X}$_{2}$\emph{B}$_{2}$\emph{T}$_{6}$ with different magnetic configurations. The parity-based Z$_{4}$ invariant is defined as~\cite{45Turner,46Ono,47Watanabe}
\begin{equation}\label{eq_1}
\begin{aligned}
  Z_{4} &= \Sigma_{\alpha=1}^{8}\Sigma_{n=1}^{n_{occ}}\frac{1+\xi_{n}(\Lambda_{\alpha})}{2}~~mod~4,
\end{aligned}
\end{equation}
where $\xi_{n}(\Lambda_{\alpha})$ is the parity eigenvalue (+1 or -1) of the \emph{n}th band at the $\alpha$th inversion-invariant momenta point $\Lambda_{\alpha}$ and $n_{occ}$ is the total number of occupied bands. Here Z$_{4}$=1 or 3 indicates a Weyl semimetallic phase, while Z$_{4}$=2 implies an axion insulator (AXI) with a quantized topological magnetoelectric effect (TME) [the axion angle $\theta=\pi$] in the case of the Chern numbers on all the 2D planes of the BZ being zeros~\cite{48Huan}. Thus, we first compute their Chern numbers in the $k_{z}=0$ and $k_{z}=\pi$ planes of these seven materials and find that the Chern numbers in both planes equal to 0. Then, we calculate the number of occupied bands with even and odd parity eigenvalues ($n_{occ}^{+}$,$n_{occ}^{-}$) at the eight inversion-invariant momenta points (only four of them are distinct) for these seven materials in their respective magnetic ground states. As shown in Table \ref{tab:I}, we see that all the invariants Z$_{4}$ equal to 2 for these seven materials, suggesting that these intrinsic FM \emph{MnGe}$_{2}$\emph{Sb}$_{2}$\emph{Te}$_{6}$, \emph{MnGe}$_{2}$\emph{Bi}$_{2}$\emph{Te}$_{6}$, and \emph{MnPb}$_{2}$\emph{Bi}$_{2}$\emph{Te}$_{6}$ as well as A-type AFM MnGe$_{2}$Sb$_{2}$Se$_{6}$, MnGe$_{2}$Bi$_{2}$Se$_{6}$, MnSn$_{2}$Sb$_{2}$Te$_{6}$, and MnSn$_{2}$Bi$_{2}$Te$_{6}$ bulk materials are all AXIs with the quantized TME.

As mentioned above, the energy differences between the FM state and the A-type AFM magnetic ground state for MnGe$_{2}$Sb$_{2}$Se$_{6}$, MnGe$_{2}$Bi$_{2}$Se$_{6}$, MnSn$_{2}$Sb$_{2}$Te$_{6}$, and MnSn$_{2}$Bi$_{2}$Te$_{6}$ are very small (about 0.02-0.05~meV/Mn), hence a small external magnetic field is sufficient to induce an FM state. The band structures of MnSn$_{2}$Sb$_{2}$Te$_{6}$, MnGe$_{2}$Sb$_{2}$Se$_{6}$, MnGe$_{2}$Bi$_{2}$Se$_{6}$, and MnSn$_{2}$Bi$_{2}$Te$_{6}$ bulk materials in the FM states with spins aligned along the \emph{c}-axis direction calculated with the inclusion of SOC are shown in Fig.~\ref{fig_SOCband}(c) and Fig. S4 in the SM, respectively. One can see that their energy bands are inverted at the $\Gamma$ point around the Fermi level. Further, we calculate the number of occupied bands with even and odd parity eigenvalues and obtain their invariants Z$_{4}$ still being 2 (see Table SII in the SM). This indicates that all the seven materials are robust AXIs regardless of being ferromagnetic or antiferromagnetic.

\begin{figure}[!th]
	\centering
	\includegraphics[width=0.47\textwidth]{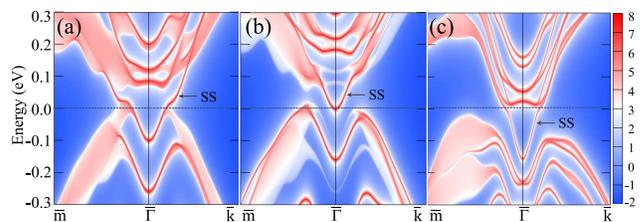}
	\caption{(Color online) Surface energy bands of the semi-infinite (111) surface of the FM \emph{MnGe}$_{2}$\emph{Sb}$_{2}$\emph{Te}$_{6}$ with onsite energy of (a) 0~eV and (b) 0.1~eV and (c) the A-type AFM MnSn$_{2}$Sb$_{2}$Te$_{6}$ bulk with onsite energy of 0~eV. Here the black arrows mark the surface states (SSs).}
	\label{fig_surfaces}
\end{figure}

Note that an AXI can be regarded as a three-dimensional gapped system with the pseudoscalar axion angle $\theta=\pi$ and a gapped surface state (SS) at the same time~\cite{49Zhao}, where an additional term $S_{\theta}=\frac{\theta}{2\pi}\frac{e^{2}}{h}\int{d^{3}xdt\emph{\textbf{E}}\cdot\emph{\textbf{{B}}}}$ related to $\theta$ is used to describe the TME. Here \textbf{\emph{E}} and \textbf{\emph{B}} are the electric and magnetic fields respectively, $e$ is the charge of an electron, and $h$ is Planck's constant. Now, we have demonstrated that these seven materials all have axion angle $\theta=\pi$. Next, we take the FM \emph{MnGe}$_{2}$\emph{Sb}$_{2}$\emph{Te}$_{6}$ and A-type AFM MnSn$_{2}$Sb$_{2}$Te$_{6}$ as examples to show that they both have gapped surface states (SSs). Since these materials are formed by stacking the eleven-layer building blocks along the \emph{c} axis via the vdW interaction, the (111) plane of their primitive cells (namely, the \emph{a}-\emph{b} plane within the vdW gap [see Fig.~\ref{fig_structure}(d)]) is their natural cleavage surface. Thus, we calculated the surface states of this plane for the FM \emph{MnGe}$_{2}$\emph{Sb}$_{2}$\emph{Te}$_{6}$ and A-type AFM MnSn$_{2}$Sb$_{2}$Te$_{6}$, as shown in Figs.~\ref{fig_surfaces}(a-b) and~\ref{fig_surfaces}(c), respectively. One sees that the A-type AFM MnSn$_{2}$Sb$_{2}$Te$_{6}$ does have the gapped SSs [Fig.~\ref{fig_surfaces}(c)]. The gap is caused by the interaction between the electrons in the SSs and the spontaneous magnetization~\cite{50Tokura}, which is similar to the case of the surface states of  MnBi$_{2}$Te$_{4}$~\cite{5Zhang,8Hao}. In comparison, the FM \emph{MnGe}$_{2}$\emph{Sb}$_{2}$\emph{Te}$_{6}$ appears to have a gapless SS [Fig.~\ref{fig_surfaces}(a)], but a careful inspection shows that the gapless SS is not protected by any symmetry.  Therefore, a small perturbation will cause it to open an energy gap. We add an onsite energy of 0.1 eV on its surface and find that it does evolve into a gapped SS [see Fig.~\ref{fig_surfaces}(b)].

To gap all the surface states and observe the TME, for the above AFM AXIs, one can apply a small in-plane magnetic field or synthesize a sample without \emph{S} symmetry~\cite{5Zhang}, where the $S=\emph{T}\tau_{1/2}$ (\emph{T} is the time-reversal symmetry and $\tau_{1/2}$ is the half translation operator). For the above FM AXIs, the interference of the gapless hinge modes on the TME can be avoided by designing a sample with a specific trigonal prism geometry to localize its chiral hinge modes on their top surface~\cite{51Liu}. Therefore, the Mn\emph{X}$_{2}$\emph{B}$_{2}$\emph{T}$_{6}$ family provides a series of feasible platform for the long-sought quantized TME.

$Discussion$: Here, we would like to emphasize the importance and advantages of our current work. First, we break through the framework based on the Bi$_{2}$Te$_{3}$ family as the parent material of the intrinsic magnetic TI, and propose a series of intrinsic FM AXIs (such as \emph{MnGe}$_{2}$\emph{Sb}$_{2}$\emph{Te}$_{6}$, \emph{MnGe}$_{2}$\emph{Bi}$_{2}$\emph{Te}$_{6}$, and \emph{MnPb}$_{2}$\emph{Bi}$_{2}$\emph{Te}$_{6}$) obtained from the \emph{X}$_{2}$\emph{B}$_{2}$\emph{T}$_{5}$ family as their parent material. Second, the FM \emph{MnGe}$_{2}$\emph{Sb}$_{2}$\emph{Te}$_{6}$, \emph{MnGe}$_{2}$\emph{Bi}$_{2}$\emph{Te}$_{6}$, and \emph{MnPb}$_{2}$\emph{Bi}$_{2}$\emph{Te}$_{6}$ films compared with the A-type AFM MnBi$_{2}$Te$_{4}$ film can achieve a comparable quantization temperature under the zero magnetic field without the constraint of the number of layers, for which we will discuss in detail in a follow-up work. Third, the Mn\emph{X}$_{2}$\emph{B}$_{2}$\emph{T}$_{6}$ family can also be extended to a large class of Mn\emph{X}$_{2}$\emph{B}$_{2}$\emph{T}$_{6}$/(\emph{B}$_{2}$\emph{T}$_{3}$)$_{n}$ family similar to MnBi$_{2}$Te$_{4}$/(Bi$_{2}$Te$_{3}$)$_{n}$ family, and Mn can also be replaced with other transition-metal or rare-earth element, thereby it provides a broad platform for studying various magnetic topological states. Finally, we offer a new design scheme for the realization of the intrinsic AXIs, and the idea may also apply to other inversion-preserved vdW TIs. Thus, we look forward to the experimental synthesis of Mn\emph{X}$_{2}$\emph{B}$_{2}$\emph{T}$_{6}$ family materials in the near future.

$Summary$: We design a new class of materials Mn\emph{X}$_{2}$\emph{B}$_{2}$\emph{T}$_{6}$ (\emph{X}=Ge, Sn, or Pb; \emph{B}=Sb or Bi; \emph{T}=Se or Te) based on \emph{X}$_{2}$\emph{B}$_{2}$\emph{T}$_{5}$ structures and study the topological properties of these materials by using first-principles electronic structure calculations. We find that the interlayer magnetic couplings can be well adjusted by substituting different elements (\emph{X}, \emph{B}, and \emph{T}) in these systems while maintaining their nontrivial topological properties. Our calculations specifically demonstrate that \emph{MnGe}$_{2}$\emph{Sb}$_{2}$\emph{Te}$_{6}$, \emph{MnGe}$_{2}$\emph{Bi}$_{2}$\emph{Te}$_{6}$, and \emph{MnPb}$_{2}$\emph{Bi}$_{2}$\emph{Te}$_{6}$ are ferromagnetic axion insulators (AXIs), while MnGe$_{2}$Sb$_{2}$Se$_{6}$, MnGe$_{2}$Bi$_{2}$Se$_{6}$, MnSn$_{2}$Sb$_{2}$Te$_{6}$, and MnSn$_{2}$Bi$_{2}$Te$_{6}$ are A-type antiferromagnetic AXIs. In addition, they all have an out-of-plane easy axis of magnetization. Thus the Mn\emph{X}$_{2}$\emph{B}$_{2}$\emph{T}$_{6}$ family materials may offer a promising platform for the realization of the long-sought quantized magnetoelectric effect and the ``high-temperature'' QAH effect in future experiments.

\begin{acknowledgments}

We wish to thank Weikang Wu, Ben-Chao Gong and Zhi-Da Song for helpful discussions. This work was supported by the National Key R\&D Program of China (Grants No. 2019YFA0308603 and No. 2017YFA0302903), the National Natural Science Foundation of China (Grants no. 11934020 and No. 11774424), the Beijing Natural Science Foundation (Grant No. Z200005), the CAS Interdisciplinary Innovation Team, the Fundamental Research Funds for the Central Universities and the Research Funds of Renmin University of China (Grants No. 16XNLQ01 and No. 19XNLG13). Y.G. was supported by the Outstanding Innovative Talents Cultivation Funded Programs 2021 of Renmin University of China. Computational resources were provided by the Physical Laboratory of High-Performance Computing at Renmin University of China.

\end{acknowledgments}

\end{document}